\documentclass{pnastwo}

\usepackage{graphicx}
\usepackage{txfonts}
\PassOptionsToPackage{hyphens}{url}
\usepackage{hyperref}

\DeclareMathAlphabet{\mathbfsf}{\encodingdefault}{\sfdefault}{bx}{n}

\newcommand{\etal}{\textit{et al.}}

\renewcommand{\vec}[1]{{\ensuremath\mathbf{#1}}}
\newcommand{\tens}[1]{\ensuremath\mathbfsf{#1}}
\newcommand{\grad}[1]{\vec{\nabla}{#1}}
\newcommand{\curl}[1]{\vec{\nabla}\times{#1}}
\renewcommand{\div}[1]{\vec{\nabla}\cdot{#1}}
\renewcommand{\f}[2]{\frac{#1}{#2}}

\newcommand{\diff}{d}
\newcommand{\dpart}[2]{\f{\partial {#1}}{\partial {#2}}}

\newcommand{\Rm}{\mathrm{Rm}}
\newcommand{\Pm}{\mathrm{Pm}}
\newcommand{\rhoi}{\rho_i}

\newcommand{\vthi}[0]{v_{\mathrm{th}i}}
\newcommand{\omegapi}{\omega_{\mathrm{p}i}}
\newcommand{\di}{d_i}

\newcommand{\bhat}{{\hat{\vec{b}}}}

\newcommand{\average}[1]{\left< {#1}\right>}

\begin{document}

\title{Turbulent dynamo in a collisionless plasma}

\author{Fran\c{c}ois Rincon\affil{1}{Universit\'e de Toulouse;
    UPS-OMP; IRAP; Toulouse, France}\affil{2}{CNRS; IRAP; 14, avenue
    Edouard Belin, F-31400 Toulouse, France},
Francesco Califano\affil{3}{Physics Department, University of Pisa,
  56127 Pisa, Italy},
Alexander A. Schekochihin\affil{4}{The Rudolf Peierls Centre for Theoretical Physics, University of
Oxford, 1 Keble Road, Oxford, OX1 3NP, United Kingdom}\affil{5}{Merton
College, Oxford OX1 4JD, United Kingdom}
\and
Francesco Valentini\affil{6}{Dipartimento di Fisica, Universit\'a della Calabria, I-87036 Rende (CS), Italy}
}

\contributor{Submitted to Proceedings of the National Academy of Sciences
of the United States of America}

\significancetext{While magnetic-field amplification by a dynamo
  effect converting kinetic flow energy into magnetic energy has long
  been demonstrated in conventional magnetohydrodynamic fluids,
  whether a similar effect is possible in
  more dynamically complex weakly-collisional plasmas, such as
  encountered in astrophysical objects on extragalactic scales, is not
  known. We present the first conclusive numerical evidence
  and dynamical picture of magnetic-field amplification by
  chaotic motions in a collisionless plasma. The results suggest that
  such a plasma dynamo may be a realizable physical effect in
  ``laboratory-astrophysics'' experiments, and support
  the idea that turbulent dynamos may significantly
  contribute to the magnetization of weakly-collisional
  high-energy-density astrophysical plasmas such
  as the intracluster medium of galaxy clusters.}

\maketitle

\begin{article}
\begin{abstract}
Magnetic fields pervade the entire Universe and affect the formation
and evolution of astrophysical systems from cosmological to planetary
scales. The generation and dynamical amplification of extragalactic
magnetic fields through cosmic times, up to $\mu$Gauss levels reported
in nearby galaxy clusters, near equipartition with kinetic energy of
plasma motions and on scales of at least tens of kiloparsecs, is a
major puzzle largely unconstrained by observations. 
A dynamo effect converting kinetic flow energy 
into magnetic energy is often invoked in that context, however
extragalactic plasmas are weakly collisional (as opposed to
magnetohydrodynamic fluids), and whether magnetic-field
growth and sustainment through an efficient turbulent
  dynamo instability is possible in such plasmas is not
  established. Fully kinetic numerical
simulations of the Vlasov equation in a six-dimensional phase space
necessary to answer this question have until recently remained beyond
computational capabilities. Here, we show by means of such simulations
that magnetic-field amplification via a dynamo instability
does occur in a stochastically-driven, non-relativistic subsonic
flow of initially unmagnetized collisionless plasma. We also
find that the dynamo self-accelerates and becomes entangled with
kinetic instabilities as magnetization increases. The
results suggest that such a plasma dynamo
  may be realizable in laboratory experiments, support
  the idea that intracluster medium (ICM) turbulence may have
  significantly contributed to the amplification of cluster
  magnetic fields up to near-equipartition levels on a timescale
  shorter than the Hubble time, and emphasize the crucial role of
  multiscale kinetic physics in high-energy astrophysical plasmas.
\end{abstract}


\dropcap{T}he generation, amplification and sustainment of magnetic
fields in Nature may be driven by a variety of physical processes, 
an important family of which are dynamo instabilities converting
kinetic energy of chaotic flows into magnetic energy. 
While fluid (collisional) magnetohydrodynamic (MHD) dynamos relevant
to the planetary, stellar and galactic contexts
had long been thought possible on the basis of idealized 
theoretical models of turbulent flows
\cite{kazantsev68,moffatt77,zeldovich84,childress95}, a major 
boost to their understanding was given by early pioneering 
numerical simulations of magnetic field amplification by 
3D MHD turbulence \cite{meneguzzi81}, a rare case of
proof-of-principle ``numerical discovery'' later followed 
by experimental evidence using liquid metals
\cite{monchaux07}. Dynamos in weakly-collisional 
plasmas, in spite of their potential relevance to cosmic
magnetogenesis \cite{kulsrud08,durrer13} on extragalactic scales
\cite{zweibel97,carilli02,vogt05,schekochihin05,medvedev06,ryu08,
vazza14,santoslima14,mogavero14,miniati15}, have thus far not achieved
such a milestone. Although dedicated laboratory experiments are
under development \cite{forest15,meinecke15},
the interactions between dynamos, collisionless damping
and kinetic-scale phenomena related to plasma magnetization are
poorly understood. Magnetization occurs when the field has grown 
sufficiently that the particles' Larmor radius becomes smaller than
the typical size $\ell$ of velocity fluctuations. It does not take
much field to achieve this: in the ICM, taking $\ell=1$~kiloparsec and
$\sim 10^7\,\mathrm{K}$ ion temperature, ions are magnetized for fields
above $10^{-13}$~Gauss, well below the level at which magnetic energy
reaches equipartition with kinetic energy of plasma motions. Past
this stage, any local change in magnetic-field strength due to compressive
or shearing motions will generate pressure anisotropies with respect
to the field by virtue of magnetic-moment conservation, rendering the
plasma everywhere unstable to fast kinetic instabilities
\cite{schekochihin05}, with potentially critical implications 
for magnetic growth and dynamics
\cite{santoslima14,mogavero14,schekochihin08,schoeffler11,kunz14,riquelme15,sironi15,hellinger15,rincon15}. 
Demonstrating whether collisionless plasma dynamos exist and how they
work requires solving Vlasov-Maxwell equations in three physical space 
dimensions (3D, this stems from antidynamo theorems in 2D
\cite{childress95}), plus three velocity space dimensions (3V). 
We performed the first numerical simulations of this
problem, which show that magnetic-field amplification 
through a turbulent collisionless plasma dynamo occurs in both
unmagnetized and magnetized regimes.

\section{Problem formulation}
Our model (see Material and methods section) describes the coupled
evolution of a quasi-neutral, non-relativistic plasma of collisionless protons 
(mass $m_i$), isothermal, fluid electrons of negligible inertia, and
electromagnetic fields $\vec{E}$ and $\vec{B}$. It formally
retains magnetic advection and induction, resistivity and the Hall
effect but, because of the isothermal-electrons assumption, does not
allow for a Biermann battery \cite{biermann50} or Weibel-like
instabilities \cite{medvedev06,weibel59} (these effects only generate very 
small seed fields and are not themselves viable dynamos). 
The equations are solved with a 3D-3V Eulerian numerical code, in a
periodic cubic spatial domain of size
$L=2000\,\pi\, d_i$, where $\di$ is the ion inertial
length, and a velocity-space range of $\pm 5$ ion thermal speeds
$\vthi$. The system is initialized with a Maxwellian ion distribution
function of uniform density $n_\mathrm{i0}$ and temperature
$T_i=m_i\vthi^2/2$, electron temperature $T_e=T_i$, and a magnetic
seed in the wavenumber range $[2\pi/L,4\pi/L]$.  The field strength,
characterized by the inverse of
$\beta=8\pi\,n_\mathrm{i0}T_i/B_\textrm{r.m.s.}^2$,
remains small enough ($\beta\gg 1$) that the Hall effect is 
negligible, but the plasma can self-magnetize if the ion Larmor
radius $\rhoi=\sqrt{\beta}\, \di$ becomes smaller than $L$. 
An incompressible, non-helical, stochastic external force
drives a plasma flow by injecting ion momentum at system scale
($k_f=2\pi/L$) with a prescribed average power density
$\varepsilon =3\times 10^{-5}\,n_{i0}m_i\vthi^3/\di$. In unmagnetized
regimes, the plasma is effectively very viscous due to the phase
mixing of momentum
by streaming ions (Landau damping), so the driving generates a
smooth (forcing-scale) chaotic, subsonic, finite-amplitude flow with a
correlation time $(k_f\vthi)^{-1}$, smaller by a factor of
Mach number than the turnover time $(k_f u_{\textrm{r.m.s.}})^{-1}$, 
where $u_{\textrm{r.m.s.}}$ is the characteristic mean ion flow
velocity. This differs from fluid dynamo simulations, in which
these timescales are comparable, and fast, inertial eddies develop
down to viscous scales. The spatial and velocity-space numerical resolution
$64^3\times 51^3$ is close to the current affordable maximum, as
characterizing the dynamo requires several simulations which must be
integrated for several turnover times with small timesteps to capture
fast kinetic physics. 

\section{Unmagnetized regime}
\begin{figure}
\includegraphics{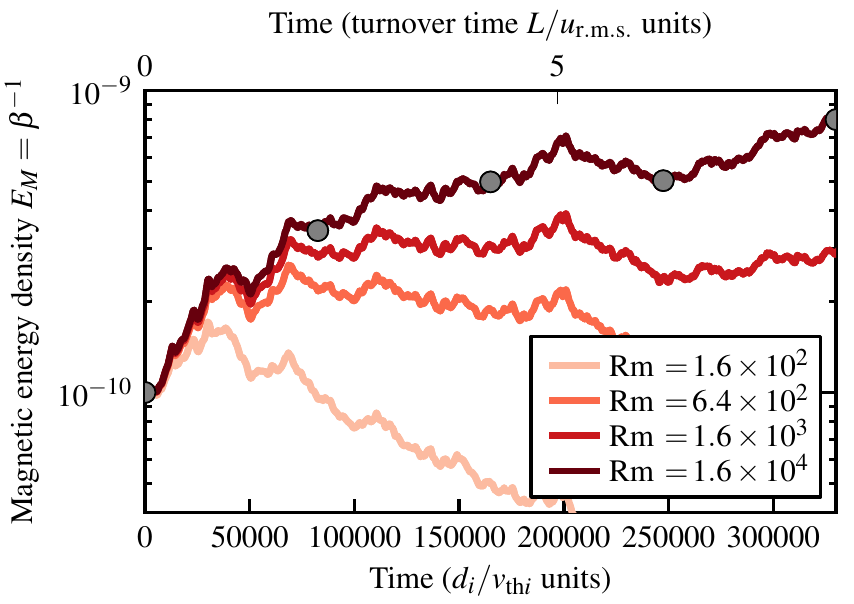}
\caption{\label{fig1}Adimensionalized magnetic-energy density
  $E_M=\beta^{-1}$ in forced simulations with decreasing
  magnetic diffusivities. The flow history is identical for all
  simulations. $L=2000\,\pi\,\di$, $k_f=2\pi/L$, $\varepsilon=3\times
  10^{-5}\,n_{i0}m_i\vthi^3/\di$, $u_\mathrm{r.m.s.}/\vthi\simeq 0.17$, and $\beta(t=0)=10^{10}$.}
\end{figure}

\begin{figure}
 \includegraphics{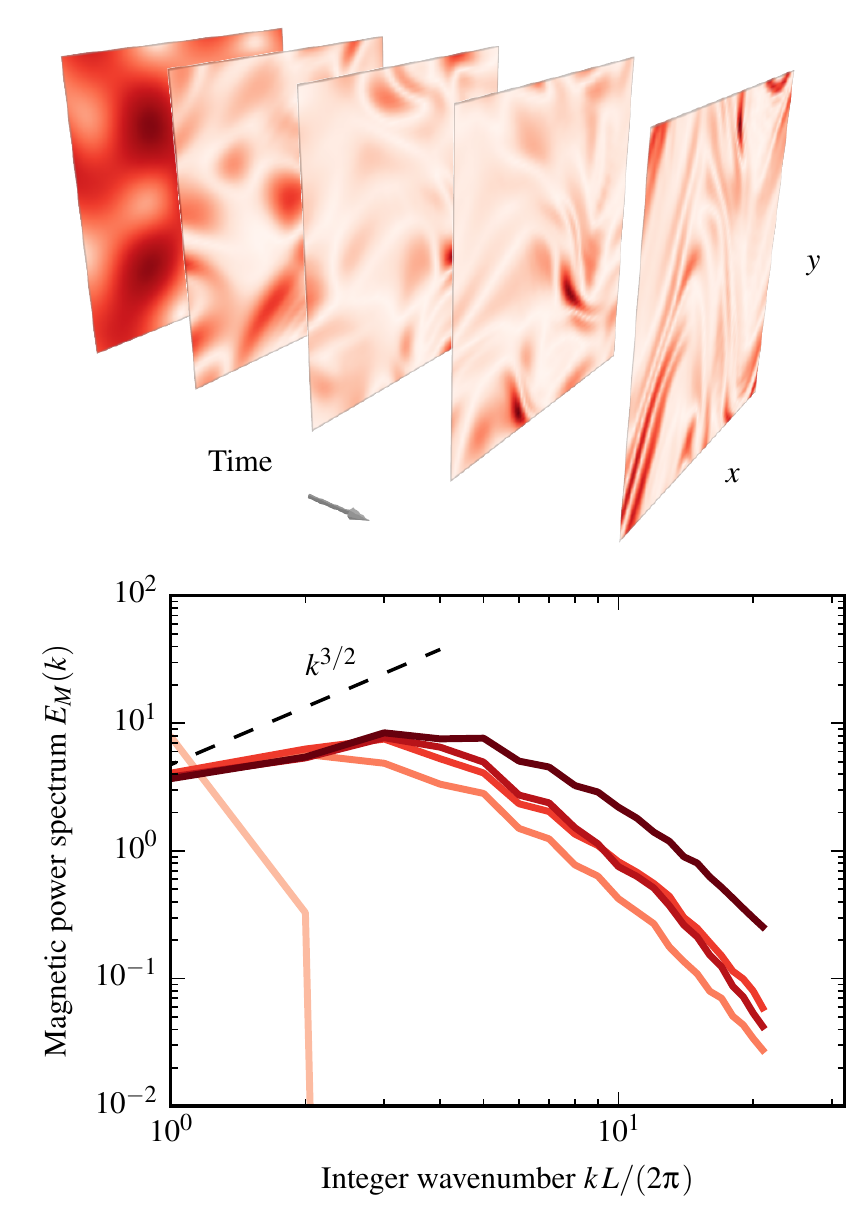}
\caption{Top: cross-sections of $|B|$ at increasing times 
  (grey circles in Fig.~\ref{fig1}) in the $\Rm\simeq 16000$
  simulation (darker regions correspond to stronger fields, the
  colormap is clipped to the amplitude range of each snapshot to
  highlight magnetic structures). Bottom: corresponding magnetic
  spectra (darker lines encode increasing times).\label{fig2}}
\end{figure}

Figure~\ref{fig1} shows the evolution of
magnetic energy in four unmagnetized simulations (initial $\beta=10^{10}$,
$\rhoi/L\simeq 16$, $u_\mathrm{r.m.s.}/\vthi\simeq 0.17$) with
magnetic diffusivities  $\eta\in[0.01,1]\,\vthi\,\di$
(magnetic Reynolds numbers $\Rm=u_\mathrm{r.m.s.}/(k_f\eta)\in
[160,16000]$)  and all other parameters fixed. The results are a
conclusive demonstration of plasma dynamo, with sustained magnetic
growth  only occurring above a critical $\Rm\simeq 1600$ at this value
of $\beta$. The magnetic energy growth rate is $\gamma\simeq 0.16\,
(u_\mathrm{r.m.s.}/L)$ for the largest $\Rm$ considered.
Figure~\ref{fig2} (see also supplementary movie 1) shows
two-dimensional snapshots of magnetic-field strength and the
corresponding energy spectra for the growing case. The field is
stretched chaotically by velocity fluctuations and develops a
characteristic folded structure with reversals perpendicular to the
field at the resistive scale $\ell_\eta$ \cite{schekochihin04}. The
trend of the spectral evolution is consistent with the formation of a
$k^{3/2}$ Kazantsev energy spectrum \cite{kazantsev68} down to
$k\sim\ell_\eta^{-1}$. These results are reminiscent of a
large-magnetic-Prandtl number ''stretch and fold''
MHD dynamo \cite{zeldovich84,childress95}
($\Pm=\nu/\eta$, where $\nu$ is the kinematic viscosity).  Considering
that the collisionless plasma flow is effectively very viscous, it is
perhaps not a surprise that its dynamo action is similar to that of
a random ``Stokes flow'' \cite{schekochihin04}. However, the critical
$\Rm$ is significantly larger than in MHD. We attribute this effect to the
shorter correlation time of collisionless eddies, which limits
their capacity to stretch the field in a sustained fashion.

\section{Magnetized regime}
\begin{figure}
\includegraphics{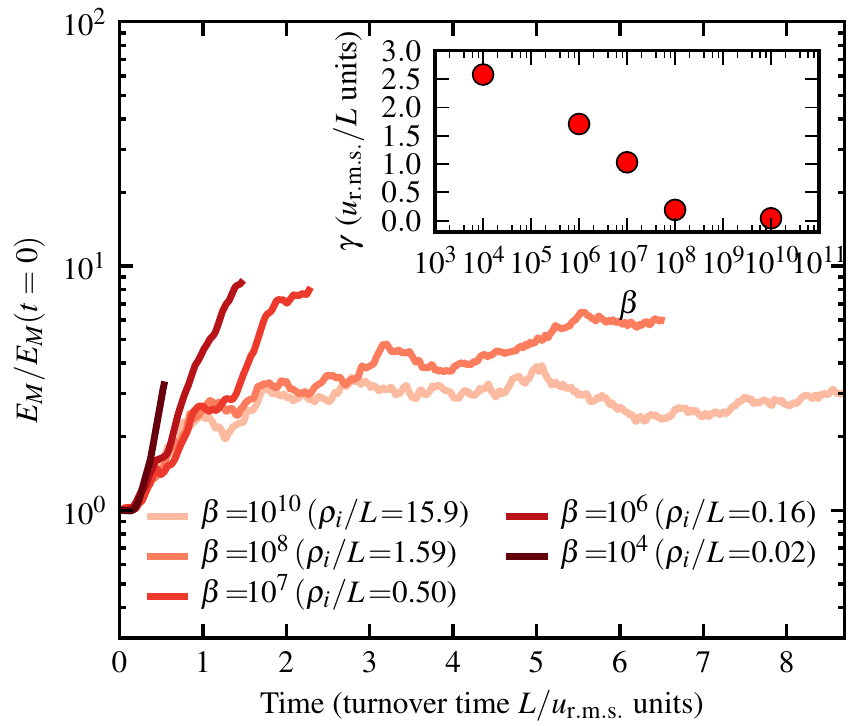}
\caption{Evolution of (normalized) magnetic-energy density in simulations
  with increasing initial magnetization (decreasing $\beta$). Inset:
  magnetic-energy growth rate versus $\beta$.
  $L=2000\,\pi\,\di$, $k_f=2\pi/L$, $\varepsilon=3\times
  10^{-5}\,n_{i0}m_i\vthi^3/\di$, $\eta=0.1\,\di\,\vthi$.\label{fig3}}
\end{figure}

\begin{figure}
\includegraphics[width=\columnwidth]{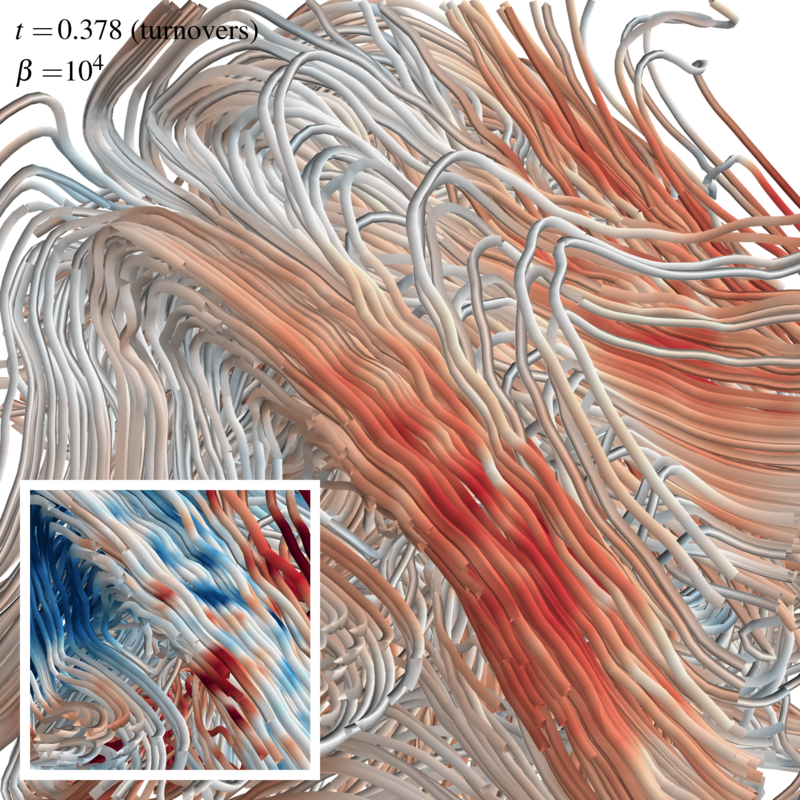}
\caption{Main image: 3D rendering of magnetic field lines subject 
  to mirror and firehose instabilities in the $\beta=10^4$
  ($\rho_i/L=0.016$) simulation (the red/blue colorscale encodes
  positive/negative ion pressure anisotropy $\Delta_i$ clipped to $\pm
  1$). Inset: close-up on field lines and scalar density
  fluctuations in the central, mirror-unstable region (the red/blue
  colorscale encodes $(n_i-n_{i0})/n_{i0}$ clipped to $\pm
  1$).\label{fig4}}
\end{figure}

As the dynamo demonstrated above proceeds, it will take the plasma
from an unmagnetized to a magnetized regime. Covering the full transition
is currently computationally prohibitive, as it requires integrating
the 3D-3V kinetic system over many (system-scale) fluid turnover times. We
instead investigated how magnetization affects magnetic growth
using several shorter simulations initialized with seed fields of
identical spatial form but different strengths ($\beta$ $\in$
$[10^{4},10^{10}]$), using the same power
input and $\eta=0.1\,\vthi\,\di$ as in the marginally stable
$\Rm\simeq 1600$, $\beta=10^{10}$ case.
Figure~\ref{fig3} shows that the magnetic-energy growth rate
increases markedly with decreasing $\beta$, leading to the conclusion
that magnetic growth is self-accelerating, and by implication that the
critical Rm is lower at lower $\beta$ (stronger seed field).
We observe a transition between a ``fluid-like'' inductive
regime, and a mixed fluid-kinetic growth regime. Figure~\ref{fig4}
(see also supplementary movie 2) shows that ion pressure anisotropies
$\Delta_i=(P_{\perp,i}-P_{\parallel,i})/P_{\perp,i}$ develop with
respect to the local field 
in the most magnetized $\beta=10^4$ ($\rho_i/L=0.016$) case
($\tens{P}_i$ is the ion pressure tensor,
$P_{\parallel,i}=\bhat\bhat:\tens{P}_i$,
$P_{\perp,i}=1/2\,(\mathrm{Tr}\,\tens{P}_i-P_{\parallel,i})$,
$\bhat=\vec{B}/|B|$). Magnetic field lines develop an angular shape in
strong field-curvature regions of negative $\Delta_i$, a nonlinear
signature of the firehose instability,
while bubbly mirror-like fluctuations are excited in regions of
positive $\Delta_i$ where the field is stretched (small-scale magnetic
depressions are associated with overdensities, see inset
in Fig.~\ref{fig4}). Relaxation of $\Delta_i$ is observed at later times in
both regions. These results are consistent with theoretical expectations
\cite{schekochihin05,schekochihin08,kunz14,riquelme15,rincon15}
and with a scenario in which magnetization and kinetic-scale
fluctuations, by impeding the free streaming of ions, enhancing
particle scattering and regulating pressure anisotropy, result in more
vigorous turbulent field amplification through an effective reduction
of flow viscosity \cite{mogavero14}. Higher-resolution simulations
with longer integration times than can be afforded currently are
required to investigate growth in this regime quantitatively.

\section{Discussion}
This paper offers a conclusive proof-of-principle demonstration 
that turbulent collisionless plasma dynamo is possible. 
This effect involves generic plasma processes independent of any
particular geometric configuration and may thus be realizable in 
``laboratory-astrophysics'' plasma experiments, provided they can
achieve sufficiently weak collisionality. Numerical evidence that the
dynamo becomes entangled with kinetic-scale dynamical phenomena as the
plasma self-magnetizes strongly suggests that future models of
weakly-collisional, magnetized turbulence in high-energy astrophysical
plasmas should at least include an effective treatment of such
multiscale interactions. For the time being, and while reconstructing the
detailed history of cosmological magnetic fields remains out of reach
observationally and computationally, our results provide a
firmer physical basis for the idea that extragalactic plasma
turbulence may significantly contribute to the amplification of seed
cosmological fields up to dynamical levels on cosmologically short
times, despite such plasmas not being simple collisional MHD
fluids. The typical magnetic-field amplification timescale
in the unmagnetized regime is an appreciable
fraction of the eddy turnover time, and our results suggest that the
dynamo self-accelerates as magnetization takes place. In the turbulent
ICM where the turnover time is believed to be no longer than $10^7$
years, probably much shorter \cite{mogavero14}, such a dynamo could
therefore in principle bring magnetic fields from typical
$10^{-21}-10^{-9}$ (at most) Gauss seed field magnitudes
\cite{kulsrud08,durrer13,medvedev06,ryu08} to $\mu$Gauss dynamical
levels in less than a Hubble time.

New supercomputing and experimental facilities should soon make it
possible to determine the parameter dependence and saturation
properties of this turbulent dynamo and to further assess its
relevance to the coevolutions of cosmic magnetic fields and
large-scale accreting structures, which are also set to be thoroughly
investigated by next-generation X-ray and radio observatories.

\begin{materials}
\section{Hybrid kinetic system} 
We consider a forced, non-relativistic, quasi-neutral
hybrid Vlasov-Maxwell system describing the coupled
evolution of collisionless protons (mass $m_i$, charge $e$), fluid,
isothermal electrons of temperature $T_e$ and negligible inertia, and
electromagnetic fields  $\vec{E}(\vec{r},t)$ and $\vec{B}(\vec{r},t)$
($\vec{r}$ and $\vec{v}$ are the spatial and velocity space
coordinates). The ion distribution function $f_i(\vec{r},\vec{v},t)$
is governed by the Vlasov equation
\begin{equation*}
\label{eq:vlasov}
\dpart{f_i}{t}+\vec{v}\cdot\grad{f_i}+\left[\f{e}{m_i}\left(\vec{E}+\f{\vec{v}\times\vec{B}}{c}\right)+\f{\vec{F}}{m_i}\right]\cdot\dpart{f_i}{\vec{v}}=0~,
\end{equation*}
where $\vec{F}(\vec{r},t)$ is an external force described below. 
The ion number density is $n_i(\vec{r},t)=\int f_i(\vec{r},\vec{v},t)\,d^3\vec{v}$, 
the mean ``fluid'' ion velocity is 
$\vec{u}_i(\vec{r},t)=\int\vec{v} f_i(\vec{r},\vec{v},t)\,d^3\vec{v}/n_i$, 
and the ion pressure tensor is
$\tens{P}_i(\vec{r},t)=m_i\int(\vec{v}-\vec{u}_i)(\vec{v}-\vec{u}_i)f_i(\vec{r},\vec{v},t)\,d^3\vec{v}$.
The electron number density $n_e$ is equal to $n_i$ at all times by
quasi-neutrality. The magnetic field evolution is governed by Faraday's equation,
\begin{equation*}
 \dpart{\vec{B}}{t}=-c\,\curl{\vec{E}}~,
\end{equation*}
and $\div{\vec{B}}=0$. The electric field is calculated from Ohm's
law, 
\begin{equation*}
\label{eq:ohm}
  \vec{E}=-\f{T_e\grad{n_e}}{en_e}-\f{\vec{u}_e\times\vec{B}}{c}+\f{4\pi\eta}{c^2}\vec{j}~,
\end{equation*}
where $\vec{j}=(c/4\pi)\,\curl{\vec{B}}$
is the current density, 
$\vec{u}_e=\vec{u}_i -\vec{j}/(e n_e)$ is the mean
electron velocity, and  $\eta$ is a uniform magnetic diffusivity.
The equations are adimensionalized using the initially uniform 
ion density $n_\mathrm{i0}$ as a reference density, the ion inertial
length $\di=c/\omegapi$, as a length scale ($\omegapi^2=4\pi\,
n_{\mathrm{i0}}\,e^2/m_i$), and $\di/\vthi$ as a timescale. 
$\vec{B}$ is expressed in units of $\vthi\sqrt{4\pi\,n_{i0}\,m_i}$,
and $\vec{E}$ in units of $\vthi^2\sqrt{4\pi\,n_{i0}\,m_i}/c$.
The adimensional magnetic energy density is the inverse of the 
plasma $\beta$

\section{Numerics} The problem is solved numerically with a
3D-3V Eulerian Vlasov code \cite{valentini07} parallelized on 1024
cores. The resistive term in Ohm's law is only included in Faraday's
equation to ensure that dynamo modes are numerically resolved.

\section{Stochastic ion momentum forcing}
An incompressible, non-helical, delta-correlated in time vector force
$\vec{F}(\vec{r},t)$ injecting ion momentum with a prescribed
statistical power density $\varepsilon$ is included in the numerical
formulation of the ion Vlasov equation using a numerical technique
borrowed from hydrodynamics \cite{alvelius99}. 
Defining the correlation tensor of the spatial Fourier transform
of the force as
\begin{equation*}
   \label{eq:forcecorrel}
\average{F_{\vec{k},i}(t)F_{\vec{k},j}^*(t')}=
\chi(k)\,\delta(t-t')\left(\delta_{ij}-k_ik_j/k^2\right),
\end{equation*}
where brackets denote ensemble averaging, 
it can be shown analytically that the (linear) response to this forcing
in unmagnetized, collisionless regimes is a time-dependent flow
$\vec{u}(\vec{r},t)$ whose correlation tensor is
\begin{equation*}
  \average{u_{\vec{k},i}(t)u^*_{\vec{k},j}(t')}= \f{\chi(k)}{8\pi
    k^2}\left(\delta_{ij}-\f{k_ik_j}{k^2}\right)\int_{-\infty}^{\infty}\!\!\!\!\!\!\diff\omega\,e^{-i\omega(t-t')}\left|Z\left(\f{\omega}{k\vthi}\right)\right|^2,
\end{equation*} 
where $Z(\zeta)$ is the plasma dispersion function \cite{fried61}.
For the forcing parameters considered here, we checked that 
the correlation tensor of the actual subsonic, chaotic flow driven at
$k=k_f$ in unmagnetized simulations is of this form to a very good
approximation, with an effective correlation time $(k_f\vthi)^{-1}$.
\end{materials}

 \appendix[Movies]

 \begin{figure}
 \caption{Movie 1:
(\protect\url{http://userpages.irap.omp.eu/~frincon/vlasov_dynamo/movie1.mp4}):
 Three-dimensional animated rendering of the dynamical
   evolution of magnetic field lines and magnetic-field strength in the
   unmagnetized $\beta=10^{10}$ dynamo simulation of Fig.~\ref{fig2}
   (darker red regions correspond to stronger fields. Unlike in
   Fig.~\ref{fig2}, the same colorscale is used at all times to
   highlight the overall growth of magnetic energy and its localization).}
 \end{figure}

 \begin{figure}
 \caption{Movie 2:
(\protect\url{http://userpages.irap.omp.eu/~frincon/vlasov_dynamo/movie2.mp4}):
  Three-dimensional animated rendering of the
   dynamical evolution of magnetic field lines and local 
   ion pressure anisotropy $\Delta_i$ in the magnetized
   $\beta=10^4$ ($\rho_i/L=0.016$) simulation. Magnetic field lines
   develop an angular shape in strong field-curvature regions of
   negative $\Delta_i$ (blue), a nonlinear signature of the firehose
   instability, while bubbly mirror-like fluctuations are excited in
   regions of positive $\Delta_i$ (red) where the field is stretched.}
 \end{figure}

\begin{acknowledgments} 
The authors thank S.~C. Cowley and
M.~W. Kunz for many valuable discussions and suggestions,
and C. Cavazzoni (CINECA, Italy) for his essential contribution 
to the code parallelisation and performance. This work was 
granted access to the HPC resources of IDRIS under the allocation
2015-i2015047188 made by GENCI, and of CINECA under
ISCRA-B allocation COLDYN.  
\end{acknowledgments}

\end{article}


\begin{thebibliography}{99}
\bibitem{kazantsev68}
Kazantsev, AP (1968) 
Enhancement of a magnetic field by a conducting fluid.
\textit{Soviet Phys JETP} 26:1031.

\bibitem{moffatt77}
Moffatt, HK (1977) \textit{Magnetic Field Generation In Electrically Conducting
  fluids}. Cambridge University Press.

\bibitem{zeldovich84}
Zel'dovich, YB, Ruzmaikin, AA, Molchanov, SA, Sokoloff,
DA (1984) Kinematic dynamo problem in a linear velocity field.
\textit{J Fluid Mech} 144:1.

\bibitem{childress95}
Childress, S \& Gilbert, AD (1995)
  \textit{Stretch, Twist, Fold. The Fast Dynamo}. Springer-Verlag.

\bibitem{meneguzzi81}
Meneguzzi M, Frisch, U \& Pouquet, A (1981) Helical and nonhelical
turbulent dynamos. \textit{Phys Rev Lett} 47:1060.


\bibitem{monchaux07}
Monchaux, R \etal\ Generation of a magnetic field by dynamo action in
a turbulent flow of liquid sodium. \textit{Phys Rev Lett} \textbf{98},
044502 (2007).

\bibitem{kulsrud08}
Kulsrud, RM \& Zweibel, EG (2008) On the origin of cosmic magnetic
fields. \textit{Rep Prog Phys} 71:046901.


\bibitem{durrer13}
Durrer, D \& Neronov, A (2013) Cosmological magnetic fields: their
generation, evolution and observation. \textit{Astron. \&
  Astrophys Rev} 21:62.


\bibitem{zweibel97}
Zweibel, EG \& Heiles, C  (1997)
Magnetic fields in galaxies and beyond. \textit{Nature} 385:131.

\bibitem{carilli02}
Carilli, CL \& Taylor, GB (2002) Cluster magnetic fields.
\textit{Annu Rev Astron Astr} 40:319.

\bibitem{vogt05}
Vogt, C \& En{\ss}lin, TA  (2005)
A Bayesian view on Faraday rotation maps -- Seeing the magnetic power
spectra in galaxy clusters.
\textit{Astron Astrophys} 434:67.

\bibitem{schekochihin05}
Schekochihin, AA, Cowley, SC, Kulsrud, RM, Hammett, GW \&
Sharma, P (2005) Plasma instabilities and magnetic field growth in
clusters of galaxies. \textit{Astrophys J} 629:139.


\bibitem{medvedev06}
Medvedev, MV, Silva, LO \& Kamionkowski, M (2006)
Cluster magnetic fields from large-scale structure and galaxy cluster
shocks. \textit{Astrophys J Lett} 642:L1.

\bibitem{ryu08}
Ryu, D, Kang, H, Cho, J \& Das, S (2008) Turbulence and magnetic
fields in the large-scale structure of the universe. \textit{Science}
320:909.

\bibitem{vazza14}
Vazza, F, Br{\"u}ggen, M, Gheller, C \& Wang, P (2014)
On the amplification of magnetic fields in cosmic filaments and galaxy clusters.
\textit{Mon Not R Astron Soc} 445:3706.

\bibitem{santoslima14}
Santos-Lima, R \etal (2014)
Magnetic field amplification and evolution in turbulent collisionless
magnetohydrodynamics: an application to the intracluster medium.
 \textit{Astrophys J} 781:84.

\bibitem{mogavero14}
Mogavero, F \& Schekochihin, AA (2014)
Models of magnetic field evolution and effective viscosity in weakly
collisional extragalactic plasmas.
\textit{Mon Not R Astron Soc} 440:3226.

\bibitem{miniati15}
Miniati, F and Beresnyak, A (2015) Self-similar energetics in large
clusters of galaxies. \textit{Nature} 523:59.

\bibitem{forest15}
Forest, CB \etal\ (2015) The Wisconsin Plasma Astrophysics
Laboratory. \textit{J Plasma Phys} 81:345810501.

\bibitem{meinecke15}
 Meinecke, J \etal\  (2015) Developed turbulence and nonlinear
 amplification of magnetic fields in laboratory and astrophysical
 plasmas. \textit{Proc Natl Acad Sci USA} 112:8211.

\bibitem{schekochihin08}
Schekochihin, AA, Cowley, SC, Kulsrud, RM,  Rosin, MS \&
Heinemann, TH (2008) Nonlinear growth of firehose and mirror fluctuations
in astrophysical plasmas. \textit{Phys Rev Lett} 100:081301.

\bibitem{schoeffler11}
Schoeffler, K, Drake, JF \& Swisdak, M (2011)
The effects of plasma beta and anisotropy instabilities on the
dynamics of reconnecting magnetic fields in the
heliosheath. \textit{Astrophys J} 743:70.

\bibitem{kunz14} 
Kunz, MW, Schekochihin, AA \& Stone, JM (2014)
Firehose and mirror instabilities in a collisionless shearing plasma.
\textit{Phys Rev Lett} 112:205003.

\bibitem{riquelme15}
Riquelme, MA, Quataert, E \& Verscharen, D (2015)
Particle-in-cell simulations of continuously driven mirror and ion
cyclotron instabilities in high beta astrophysical and heliospheric
plasmas.
\textit{Astrophys J} 800:27.

\bibitem{sironi15}
{Sironi}, L \& {Narayan}, R (2015) Electron Heating by the Ion Cyclotron
Instability in Collisionless Accretion Flows. I. Compression-driven
Instabilities and the Electron Heating
Mechanism. \textit{Astrophys J} 800:88.

\bibitem{hellinger15}
Hellinger, P \& {Tr{\'a}vn{\'{\i}}{\v c}ek}, PM (2015) Proton
temperature-anisotropy-driven instabilities in weakly collisional
plasmas: Hybrid simulations. \textit{J Plasma Phys} 81:305810103.

\bibitem{rincon15}
Rincon, F, Schekochihin, AA \& Cowley, SC (2015)
Non-linear mirror instability.
 \textit{Mon Not R Astron Soc} 447:L45.

\bibitem{biermann50}
Biermann, L (1950)
\"Uber den ursprung der magnetfelder auf sternen
und im interstellaren raum. \textit{Z Naturf} 5:65.

\bibitem{weibel59}
Weibel, ES (1959) Spontaneously growing transverse waves in a plasma
due to an anisotropic velocity distribution.
\textit{Phys Rev Lett} 2:83.

\bibitem{schekochihin04}
 Schekochihin, AA, Cowley, SC \& Taylor SF (2004) Simulations of the
 small-scale turbulent dynamo. \textit{Astrophys J} 612:276.

\bibitem{valentini07}
Valentini, F, Tr{\'a}vn{\'\i}\v{c}ek, P, Califano, F, Hellinger,
P \& Mangeney, A (2007)
A hybrid-Vlasov model based on the current advance method for the
simulation of collisionless magnetized plasma.
 \textit{J Comp Phys} 225:753.

\bibitem{alvelius99}
Alvelius, K (1999) Random forcing of three-dimensional homogeneous
turbulence. \textit{Phys Fluids} 11:1880.

\bibitem{fried61}
Fried, BD \& Conte, SD (1961) \textit{The Plasma Dispersion Function}.
New York: Academic Press.

\end{thebibliography}
\end{document}